\documentclass[%
 reprint,
 amsmath,amssymb,
 aps,
10pt]{revtex4-1}
\usepackage{xcolor}
\usepackage{amsmath}
\usepackage{graphicx}% Include figure files
\usepackage{dcolumn}% Align table columns on decimal point
\usepackage{bm}% bold math
\newcommand{\be}{\begin{equation}}
\newcommand{\ee}{\end{equation}}
\newcommand{\bea}{\begin{eqnarray}}
\newcommand{\eea}{\end{eqnarray}}

\newcommand{\bb}{\bibitem}
 
\newcommand{\eqn}{\begin{eqnarray}}
\newcommand{\eqnx}{\end{eqnarray}}

\begin{document}

%\preprint{APS/123-QED}

\title{Inflationary Twin Models}% Force line breaks with \\

\author{C. Adam}
 \affiliation{Departamento de F\'isica de Part\'iculas, Universidad de Santiago de Compostela and Instituto Galego de F\'isica de Altas Enerxias (IGFAE) E-15782 Santiago de Compostela, Spain}
 \author{D. Varela}
\affiliation{Departamento de F\'isica de Part\'iculas, Universidad de Santiago de Compostela and Instituto Galego de F\'isica de Altas Enerxias (IGFAE) E-15782 Santiago de Compostela, Spain}

\date{\today}% It is always \today, today,

\begin{abstract}
We introduce the concept of inflationary twin models, that is, a class of generalized (kinetic or "K-inflation") field theories which lead exactly to the same cosmological evolution during the inflationary period as a given standard scalar field theory of inflation. The twin concept permits to introduce generalized, K-inflation theories in a controlled manner, maintaining some inflationary predictions unaltered. Further, this concept allows to extend analytical tools like the slow-roll expansion to the realm of K-inflation theories, facilitating their investigation. 
Twins of a standard scalar field model of inflation still may result in different results for some inflationary observables, because of their nontrivial scalar speed of sound. This implies that non-standard twins may lead to completely viable models of inflation, even if their standard twin is ruled out by observations. We provide some explicit examples of this possibility, including the physical case of a dilaton--Dirac-Born-Infeld theory. 
   
\end{abstract}

\pacs{Valid PACS appear here}% PACS, the Physics and Astronomy
\maketitle

%\tableofcontents

\section{ Introduction} 
Cosmological inflation \cite{starob1}-\cite{Baumann}, originally conceived to explain the absence of magnetic monopoles and the almost perfect spatial flatness and isotropy of the visible universe,  nowadays receives particularly strong support from the observations of the power spectra of cosmic microwave background (CMB) temperature anisotropies and their slight scale dependence \cite{WMAP1}-\cite{Planck}. Moreover, the most recent results of  \cite{Planck} are fully compatible with the simplest version of inflation, where cosmological inflation is driven by one scalar field (the "inflaton"), see \cite{enzy} for an extensive list. There are no indications for features (like non-Gaussianities in the CMB temperature correlations) which would require more elaborate versions of cosmological inflation like, e.g., multi-field inflation.  

The inflaton field $\phi$ is frequently described by a standard scalar field lagrangian
\be \label{stand-L}
\bar{\mathcal{L}} = X - V(\phi) , \quad X \equiv  \frac{1}{2} g^{\mu\nu} \partial_\mu \phi \partial_\nu \phi
\ee 
where $X$ is the standard kinetic term and $V(\phi)$ is a potential. In many models, the inflaton may take values near the Planck scale, therefore higher powers are not suppressed and the potential $V$ is, a priori, rather arbitrary.  For the same reason, higher powers of $X$ should be taken into account. Indeed, general lagrangians 
$\mathcal{ L}(X,\phi)$ have been considered and are known as K-inflation \cite{k-infl}-\cite{Kachru2007}. 
In particular, although the correct inflaton model has not yet been determined from fundamental theories like string theory or supergravity (SUGRA), they tend to give rise to rather specific generalized lagrangians. We shall consider a dilaton-DBI (=Dirac-Born-Infeld) model as a simple and relevant example.  

For standard cosmological inflation lagrangians (\ref{stand-L}), analytical methods like the slow-roll expansion are available, whereas generalized lagrangians usually have to be investigated case by case (but see \cite{Vennin3}).
It would, therefore, be desirable to introduce generalized lagrangians in a controlled way, which allows to extend the analytical methods of the standard case and maintains certain inflationary properties unaltered. This possibility is precisely realized by "inflationary twin models"
\cite{Twin-models}. 
By definition, an inflationary twin model of a standard inflaton lagrangian (\ref{stand-L}) is a generalized inflaton lagrangian $\mathcal{ L}(X,\phi)$  leading to the {\em same} cosmological scale factor $a(t)$, inflaton solution $\phi(t)$ and energy density $\rho (t)$ as the given standard theory $\bar{\mathcal{L}}$. Here, the Friedmann-Robertson-Walker (FRW) line element is
\be\label{FRW-metric}
ds^2  = dt^2 - a^2 (t) \delta_{ij} dx^i dx^j ,
\ee
and $t$ is cosmological time.
The usefulness of the twin model concept is closely related to the fact that twin models of a given model 
$\bar{\mathcal{L}}$ can be found by imposing some purely algebraic conditions on generalized lagrangians $\mathcal{ L}(X,\phi)$ \cite{twins-adam1}. Solving the FRW equations (either for $\bar{\mathcal{L}}$ or for 
$\mathcal{ L}(X,\phi)$) is not required. Twin models $\mathcal{ L}(X,\phi)$ of a standard theory 
$\bar{\mathcal{L}}$ may still lead to different predictions for some inflationary observables, because of a different speed of sound for scalar fluctuations, which affects the scalar power spectrum but not the tensor one. It is, in fact, possible to "tune" the speed of sound
and the corresponding values for inflationary observables in a controlled way. 

In the next section, we introduce the FRW equations both for  $\bar{\mathcal{L}}$ and for $\mathcal{ L}(X,\phi)$, and the algebraic twin conditions. We also consider how inflationary observables are affected. Then we study twin models of the simplest large-field inflationary model as a specific example, and compare with the most recent observational data. We use natural units such that $c=\hbar =1$, so the only dimensionful universal constant is the reduced Planck mass $M_{\rm P}$ related to Newton's constant $G$ via 
$
M_{\rm P} = (8\pi G)^{-\frac{1}{2}}.
$

%\vspace*{0.2cm}

\section{ FRW equations and twin models}
The FRW equations for the time evolution of an isotropic universe filled with a perfect fluid are
($\dot H \equiv (dH/dt)$)
\bea \label{FRW-eq1}
H^2 &=& \frac{1}{3M_{\rm P}^2}\rho (t) , \\ \label{FRW-eq2}
\dot H &=& - \frac{1}{2M_{\rm P}^2}\left( \rho (t) + p(t)\right)
\eea
where $H=\dot a/a$ is the Hubble function and $\rho$ and $p$ are the energy density and pressure of the perfect fluid. The generalized lagrangians $\mathcal{ L}(X,\phi)$ with a time-dependent inflaton $\phi (t)$ are precisely of this type, with ($\mathcal{L}_{,X} \equiv \partial_X \mathcal{L}$)
\be
\rho = 2X\mathcal{L}_{,X} -\mathcal{L}, \quad p = \mathcal{L} .
\ee
Inflation is defined as a phase of accelerated expansion, $\ddot a > 0$. Owing to the relation
\be
\frac{\ddot a}{a} = -\frac{1}{6M_{\rm P}^2}\left( \rho + 3 p\right) ,
\ee
this implies $\rho + 3 p <0$. In the case of a standard inflationary lagrangian (\ref{stand-L}) with 
\be
\bar \rho = X+V, \quad \bar p = X - V
\ee
(where now $X = \frac{1}{2} \dot \phi^2$), this implies $V>\dot \phi^2$, i.e., the potential energy dominates over the kinetic one.

Obviously, a generalized model $\mathcal{ L}(X,\phi)$ and a standard lagrangian $\bar{\mathcal{L}}$ will lead to the same inflationary evolution provided that $\rho (t) =\bar \rho (t)$ and $p(t) = \bar p(t)$, but apparently the corresponding solutions of the FRW equations are required to establish this twin property. To show that this is not the case, we need the so-called 
Hamilton-Jacobi formalism \cite{bond1990}-\cite{kinney1997}, also known as the
superpotential method \cite{nunez2004}-\cite{infl-adam1}. It consists in assuming that during inflation, i.e., while the cosmological evolution is determined entirely by the inflaton, $H$ may be treated as a function of $\phi$ instead of $t$,
\be \label{H-W}
H = \sqrt{2} M_{\rm P} W(\phi) \quad \Rightarrow \quad \dot H = \sqrt{2} M_{\rm P}W_{,\phi} \dot \phi 
\ee
where $W(\phi)$ is the superpotential (dimensionless in our conventions; for more details see \cite{infl-adam1}). For a standard Lagrangian $\bar{\mathcal{L}}$, the second FRW equation (\ref{FRW-eq2}) then leads to
\be \label{dot-H}
\dot H = -\frac{1}{2M_{\rm P}^2} \dot\phi^2 \quad \Rightarrow \quad \dot \phi = -\left( \sqrt{2}M_{\rm P}\right)^3 W_{,\phi}
\ee
which may be considered as a transformation from cosmological time $t$ to the new "time" variable $\phi$. Finally, inserting (\ref{H-W}) and (\ref{dot-H}) into the first FRW equation (\ref{FRW-eq1}), we arrive at the so-called superpotential equation,
\be \label{superpot-eq}
V = 6 M_{\rm P}^4 W^2 - 4 M_{\rm P}^6 [W_{,\phi}]^2
\ee
which, for a given potential $V$, is
a first-order ordinary differential equation (ODE) for $W$. To establish the algebraic twin conditions, we start from Eq. (\ref{dot-H}), which implies
\be
 X = 4 M_{\rm P}^6 [W_{,\phi}]^2 .
\ee
This is a first-order ODE for $\phi(t)$, but for our purposes it is more useful to interpret it as an algebraic equation in phase space $(\dot \phi ,\phi)$. Imposing this equation on a general phase space function will be called "on-shell evaluation" and denoted by a vertical line in the following, i.e., 
\be
F(X,\phi)\vert \equiv F(X= 4 M_{\rm P}^6 [W_{,\phi}]^2, \phi) .
\ee
It then follows easily that the second FRW equations for $\mathcal{ L}(X,\phi)$ and $\bar{\mathcal{L}}$ will coincide provided that $(\rho + p)\vert = (\bar \rho + \bar p)\vert$, i.e.,
\be \label{1-twin}
2X\mathcal{L}_{,X} \vert = 2X\vert = 8M_{\rm P}^6 [W_{,\phi}]^2 \quad \Rightarrow \quad 
\mathcal{L}_{,X} \vert =1,
\ee
i.e., the on-shell evaluation of $\mathcal{L}_{,X}$ coincides with the off-shell value of
$\bar{\mathcal{L}}_{,X} \equiv 1$. Finally, the first FRW equations for $\mathcal{ L}(X,\phi)$ and $\bar{\mathcal{L}}$ will coincide if
\be \label{2-twin}
\mathcal{L}\vert = \bar{\mathcal{L}} \vert = 4 M_{\rm P}^6 [W_{,\phi}]^2 - V = 
8 M_{\rm P}^6 [W_{,\phi}]^2 - 6 M_{\rm P}^4 W^2.
\ee
Frequently, the generalized lagrangian will contain a "potential" term $U(\phi)$, i.e., $\mathcal{L}(X,\phi )= F(X, \phi) - U(\phi)$, then the second twin condition (\ref{2-twin}) just serves to determine $U$ in terms of $W$ and $F(X,\phi)\vert$. 

One first simple class of examples of twins is provided by lagrangians of the type
\be \label{prod-twins}
\mathcal{L} = f(\phi) g(X) - U(\phi).
\ee
The first twin condition (\ref{1-twin}) leads to
$
f(\phi) = [g_{,X}]^{-1}\vert
$
whereas the second condition (\ref{2-twin}) gives
\be
U = \frac{g\vert}{g_{,X}\vert} + 6 M_{\rm P}^4 W^2 - 8 M_{\rm P}^6 [W_{,\phi}]^2 .
\ee
Another class of examples consists of a power series expansion in $X$,
\be \label{series-twins}
\mathcal{L} = \sum_{i=1}^n f_i (\phi) X^i - U(\phi).
\ee
Here, the first twin condition just imposes one condition
\be \label{series-first-cond}
\mathcal{L}_{,X}\vert = \sum_{i=1}^n if_i (\phi)  \left( 4M_{\rm P}^6 [W_{,\phi}]^2 \right)^{i-1} =1
\ee
on the otherwise arbitrary functions $f_i (\phi)$, whereas the second condition determines $U$ in terms of the $f_i$ and $W$. 

The brief discussion of the twin model concept and the corresponding twin conditions given in this section are, in principle, self-contained and sufficient for our purposes. As the method is not widely known, however, we re-derive the twin conditions in appendix A in a more detailed manner within the Hamilton-Jacobi formalism. We hope that this second derivation provides a more concise and transparent picture about the meaning and scope of the twin model concept.

%\vspace*{0.2cm}

\section{ Hubble flow parameters and inflationary observables} 
Up to now, our considerations permit to construct twins of arbitrary standard scalar field lagrangians (\ref{stand-L}), but now we want to restrict to the so-called "slow-roll inflation". To be of cosmological relevance, inflation has to last sufficiently long, and slow-roll inflation is the simplest way to achieve this.
For a standard field theory $\bar{\mathcal{L}}$, this means that the potential $V$ and the initial value $\phi_i = \phi (t_i)$ must be chosen such that the condition $V>\dot \phi^2$ holds long enough to produce the required amount of inflation, i.e., the field $\phi$ "slowly rolls down the potential".
Strictly speaking, this interpretation no longer holds for generalized lagrangians, but the slow-roll conditions may be expressed in terms of the Hubble function $H$, and in this form they apply equally well to a general lagrangian, because twin models lead to the same Hubble function and the same slow-roll parameters. 

One convenient set of dimensionless slow-roll parameters are the so-called Hubble flow functions $\epsilon_k$ \cite{Schwarz}, \cite{Vennin2}. Here, $\epsilon_0 = (H_i/H)$, where $H_i \equiv H(t_i)$ is the Hubble function evaluated at some initial instant $t_i$ before the onset of the cosmologically relevant phase of inflation (the precise value of $t_i$ is irrelevant for our purposes). For $k\ge 1$, the $\epsilon_k$ are defined recursively, 
\be
\epsilon_{k+1} = \frac{d}{dN} \ln \epsilon_k .
\ee 
In this definition, the number of e-folds
\be
N = \ln \left( \frac{a(t)}{a(t_i)}\right) = \ln \left( \frac{a}{a_i} \right)
\ee
 is used as a new, dimensionless "time" variable, where the different "time" variables are related via
\be \label{times}
dN = Hdt = -\frac{1}{2M_{\rm P}^2}\frac{W}{W_{,\phi}} d\phi .
\ee
Intuitively, $\epsilon_1$ measures the rate of change of $H$, $\epsilon_2$ measures the rate of change of $\epsilon_1$, etc., and the slow-roll regime corresponds to $\epsilon_k << 1 \; \forall \; k\ge 1$.

In particular, 
\be
\epsilon_1 = \frac{d}{dN} \ln \frac{H_i}{H} = - \frac{\dot H}{H^2} = 2M_{\rm P}^2 \frac{W_{,\phi}^2}{W^2}
\ee
is related to accelerated expansion via
\be
\frac{\ddot a}{a} = H^2 \left( 1 +\frac{\dot H}{H^2}  \right) \equiv H^2 (1-\epsilon_1 ).
\ee
Inflation occurs for $\epsilon_1 < 1$. $\epsilon_1 =0$, (i.e., $\dot H =0$) leads to a de Sitter universe, whereas $\epsilon_1 <0$ corresponds to phantom matter and is not possible for standard scalar field inflation.
Further, 
\bea
\epsilon_2 &=& -2M_{\rm P}^2 \frac{W_{,\phi}}{W}\frac{d}{d\phi} 2\ln \frac{W_{,\phi}}{W}
= 2\epsilon + 2\delta , \\ &&  \quad \delta \equiv -2M_{\rm P}^2 \frac{W_{,\phi\phi}}{W}, \quad 
\epsilon \equiv \epsilon_1
\eea
where $\epsilon $ and $\delta$ is another pair of frequently used slow-roll parameters. Finally, in leading order slow-roll (=lsr), we have $W_{\rm lsr} =M_{\rm P}^{-2}\sqrt{V/6}$ (\cite{footnote}) and
\bea
\epsilon_{\rm lsr}&=& \frac{M_{\rm P}^2}{2}\frac{V_{,\phi}^2}{V^2} \equiv \epsilon_V \\ \delta_{\rm lsr} &=&  -M_{\rm P}^2 \frac{V_{,\phi\phi}}{V} + \frac{M_{\rm P}^2}{2}\frac{V_{,\phi}^2}{V^2} \equiv \delta_V + \epsilon_V
\eea
where $\epsilon_V$ and $\delta_V$ are called potential slow-roll parameters.

The most relevant inflationary observables are the scalar and tensor power spectra and their scale dependence. Within the slow-roll approximation, the dimensionless scalar and tensor power spectra are \cite{k-infl2}
\bea \label{Del-s}
\Delta_s^2 (k) &=& \left. \frac{1}{8\pi^2 M_{\rm P}^2}\frac{H^2}{\epsilon \, c_s}\right|_{t'_k} ,
\\
 \label{Del-t}
\Delta^2_t (k) &=& \frac{2}{\pi^2 M_{\rm P}^2} \left. H^2 \right|_{t_k}.
\eea
Here, $\left. \right|_{t_k}$ (or $\left. \right|_{t'_k}$) indicates that the expression should be evaluated at the instant $t_k$ (or $t'_k$) when the corresponding tensor mode (or scalar mode)  with wave number $k$ crosses the horizon, defined implicitly by $k = a(t_k) H(t_k)$ (or $k= c_s^{-1}(t'_k)a(t'_k)H(t'_k)$). Further, $c_s$ is the speed of sound of scalar fluctuations, which is equal to one for a standard lagrangian $\bar{\mathcal{L}}$, but not in general,
\be \label{speed-of-sound}
c_s^2 (\phi) = \frac{p_{,X}\vert}{\rho_{,X}\vert} = \left. \left( 1+2X \frac{\mathcal{L}_{,XX}}{\mathcal{L}_{,X}} \right)^{-1} \right| 
\ee
(we remind the reader that the vertical line means on-shell evaluation, such that $c_s$ in the above equation is expressed directly as a function of $\phi$ which, however, may be related to other "time" variables via (\ref{times})). 

Owing to the different propagation velocities of scalar ($c_s$) and tensor ($c=1$) modes, their horizon crossing times $t'_k$ and $t_k$ are slightly different. They are, however, equal in lsr, because in leading order a variation of $k$ in the horizon crossing condition is induced only by the (huge) variation of $a(t)$ in the inflationary epoch, whereas the variations of $H(t)$ and $c_s (t)$ are subleading  (see also Eqs. (\ref{dk-dN}), (\ref{dk-dN'}) below).
  
We also need the tensor-to-scalar ratio
\be
r = \frac{\Delta^2_t (k)}{\Delta^2_s (k)}= 16 c_s \epsilon 
\ee
and the scalar spectral index
\be
n_s -1 \equiv k\frac{d \ln \Delta^2_s}{d k}.
\ee
With the horizon crossing conditions $k = aH\vert_{t_k} $, $k = c_s^{-1} aH\vert_{t'_k} $  and their corresponding slow-roll approximations
\bea \label{dk-dN}
\frac{1}{k} d  k  &=& ( 1 - \epsilon_1)dN \simeq dN , \\ \label{dk-dN'}
\quad \frac{1}{k} d  k  &=& ( 1 - s- \epsilon_1)dN \simeq dN
\eea
we get
\be
n_s -1 \simeq \frac{d}{dN} \ln \left( \frac{H^2}{M_{\rm P}^2 \epsilon \, c_s}\right) = -2\epsilon_1 - \epsilon_2 -s
\ee
where 
\be \label{s-index}
s = k \frac{d}{dk} c_s \simeq  \frac{d}{dN} \ln c_s 
%%= \frac{\dot c_s}{Hc_s} 
= \frac{\dot{c_s^2}}{2Hc_s^2} = -M_{\rm P}^2\frac{W_{,\phi} (c_s^2)_{,\phi}}{Wc_s^2}
\ee
measures the scale dependence of $c_s$. As a standard inflation model $\bar{\mathcal{L}}$ and its non-standard twins lead to the same Hubble flow functions $\epsilon_k$, possible differences in inflationary observables must be caused by a nontrivial speed of sound $c_s$ and its scale dependence $s$. 
%%%%%%%%%%
We want to emphasize, again, that the twin model concept allows to introduce these differences in a controlled and systematic way by an appropriate choice of the non-standard twin (we shall give some concrete examples in the next section). Indeed, all twins obeying the on-shell condition $\mathcal{L}_{,XX}\vert =0$ lead to $c_s=1$ and, therefore, exactly to the same inflationary observables as their standard twin.  Generalized models which fulfill the on-shell condition $(X\mathcal{L}_{,XX})\vert = \rm{const.}$ (remember $\mathcal{L}_{,X}\vert =1$) lead to a constant scalar speed of sound $c_s$, implying a different absolute value of the scalar power spectrum but the same scale dependence. This implies a different tensor-to-scalar ratio $r$ but the same scalar spectral index  $n_s$ like the standard twin (one interesting model realizing this possibility is the dilaton-DBI model, as we shall see in the next section). Finally, generalized models which do not fulfill these on-shell conditions will lead to completely different predictions for inflationary observables. The corresponding terms in $\mathcal{L}(X,\phi)$ may be turned on in a smooth way, allowing to "tune" these deviations from the standard predictions in a controlled fashion. 
%%%%%%%%%%%

The latest Planck measurements \cite{Planck} found the following best-fit values,
\bea
(n_s -1)(k_*) &=& - 0.035 \pm 0.004 \\
\Delta^2_s (k_*) &=& (2.099 \pm 0.028) \cdot 10^{-9} \label{mes-Del-s} \\
r (k_*) &\le & 0.070.
\eea
The scales $k_{\rm CMB}$ relevant for these measurements range from scales close to the Hubble distance of the current universe (corresponding to scales which re-entered the horizon recently), $ 1/H_0 \simeq 14 \, {\rm Gly} \simeq 4 \, {\rm Gpc}$ (Gpc = Gigaparsec, Gly = Giga light years), to about $10^{-3}/H_0$, i.e., 
$1/H_0 \ge 1/k_{\rm CMB} \ge 10^{-3}/H_0$. Further,
 $k_*$ is a pivot scale within this range, $k_* = 0.05 \, ({\rm Mpc})^{-1}$, i.e., $1/k_* = 20 \, {\rm Mpc}\simeq (1/200) (1/H_0)$. 
 
 For the tensor-to-scalar ratio $r$ only an upper bound could be established.
Further, no evidence of a scale dependence of $n_s -1$ is found in a scale range $0.005 \, {\rm Mpc}^{-1} \le k_{\rm CMB} \le 0.2 \, {\rm Mpc}^{-1}$. 
We remark for later use that within slow-roll inflation, where $dk/k \simeq dN$, this scale range is related to a range in the e-fold number of $\delta N \simeq \ln (0.2/0.005) \simeq 3.69$. 
 As $n_s -1$ is determined with a precision of about 10\%, a prediction of its running at or below the 10\% level in this scale range, therefore, should not be considered an argument to exclude the corresponding model. 

%\vspace*{0.2cm}

\section{Twins for the quadratic potential} 
As a concrete example, we will calculate inflationary observables for twins of the standard inflaton model with a quadratic potential, $V=M^2 \phi^2$. For convenience, we define the dimensionless variables
\be
\varphi = ({\phi}/{M_{\rm P}}) \; , \quad m = ({M}/{M_{\rm P}}).
\ee
Further, for this inflaton model it turns out that the inflationary observables agree at the 1\% level between the exact and the lsr Hubble flow parameters, therefore the lsr approximation is sufficient for our purposes. In lsr we have (remember that all Hubble flow functions are {\em the same}  for a standard theory and its twins)
\be
\epsilon_V = - \delta_V = ({2}/{\varphi^2})
\ee
leading to 
\be
n_s - 1 \stackrel{\rm lsr}{=} -4\epsilon_ V -s \stackrel{!}{=} -0.035 .
\ee

In general (for nonzero $s$)  the value of $\epsilon_V (\varphi_*)$ will, therefore, be different for the standard theory and its twins (here $\phi_*=M_{\rm P}\varphi_*$ is the field value at time $t_*$ where $k_* = a(t_*) H(t_*)$).
For simplicity, however, we will consider only twins where the speed of sound $c_s <1$ is almost constant (constant in lsr). We get $s_{\rm lsr} =0$ and
\be
\epsilon_V (\varphi_*)= 0.00875 \quad \Rightarrow \quad \varphi_* = 15.12.
\ee
For the tensor-to-scalar ratio we get
\bea &&
r(\varphi_*) \stackrel{\rm lsr}{=} 16\epsilon_ V(\varphi_*) c_s (\varphi_*) = 0.14 \, c_s(\varphi_*) \, \stackrel{!}{\le}\,  0.07 \\
&& \Rightarrow \; c_s(\varphi_*) \,  \le \,  0.5 \, . \label{cs-bound}
\eea 
The standard quadratic potential is, therefore, strongly disfavored by current observational data, but this not the case for its twins, where $c_s \le 0.5$ can be achieved easily.

Within lsr, inflation ends at $\epsilon_V(\varphi_e) =1 \; \Rightarrow \; \varphi_e = \sqrt{2}$. This allows us to determine the number of e-folds from $t_*$ to $t_e$ (i.e., from $\varphi_*$ to $\varphi_e$). Using $dN = -(1/\sqrt{2\epsilon}) d\varphi$ we get within lsr
\be
\Delta N_* = N(\varphi_e) - N(\varphi_*) \stackrel{\rm lsr}{=} \int_{\varphi_{e}}^{\varphi_*} \frac{\varphi d\varphi}{2} = \frac{\varphi_*^2}{4} - \frac{1}{2} = 56.65 \, .
\ee
This also permits us to estimate the running of $n_s -1$ in the relevant scale range. We have approximately $\epsilon_V(\varphi_*) \simeq (1/2\Delta N_*)$ and therefore, $(\delta \epsilon_V /\epsilon_V) \simeq -(\delta N/\Delta N_*)$, leading to
\be
\left| \frac{\delta (n_s -1)}{n_s -1} \right| \simeq \frac{\delta N}{\Delta N_*} = \frac{3.69}{56.65} \simeq 0.065
\ee
which is safely below 10\%.   Finally, from (\ref{Del-s}) and the measured value (\ref{mes-Del-s}) we find for the mass parameter $m$ in lsr
\be
m^2 \stackrel{\rm lsr}{=} 2.099\cdot 10^{-9} \frac{48 \pi^2 c_s(\varphi_*)}{\varphi_*^4} \simeq 1.90 \cdot 10^{-11} c_s (\varphi_*)
\ee
and for the energy density at $\varphi_*$ in lsr
\be
\frac{\rho (\varphi_*)}{M_{\rm P}^4} \stackrel{\rm lsr}{=} m^2 \varphi_*^2 \simeq 4.35 \cdot 10^{-9}c_s(\varphi_*) 
\ee
(remember that twins will depend on the mass parameter $m$ via the superpotential equation).

Before considering some particular examples of twins for the quadratic potential, we have to explain a simplification which occurs for this potential in lsr. Indeed, the superpotential in lsr is linear in the inflaton field in this case and, therefore, its derivative $(W_{\rm lsr})_{,\phi} = m/(\sqrt{6}M_{\rm P})$ as well as the on-shell evaluation of the kinetic term, $X\vert = (2/3) m^2 M_{\rm P}^4$, are constant. This implies that if $\mathcal{L}_{,XX}$ only depends on $X$ and not explicitly on $\phi$, then its on-shell evaluation and, consequently, the speed of sound are constant in lsr, too, and $s_{\rm lsr}=0$, see (\ref{speed-of-sound}), (\ref{s-index}). In particular, all twin models of the product form (\ref{prod-twins}) are of this type. Indeed, we calculate easily $\mathcal{L}_{,XX} = (g_{,XX}/g_{,X}\vert)$ where $g(X)$ only depends on $X$, by assumption, and, therefore, $\mathcal{L}_{,XX} \vert \simeq$ const. in lsr.

The simplest twin in this class is given by a $g$ which is just a higher power of $X$, 
\be \label{hi-pow}
g(X)=X(X/\bar M^4)^{n-1}.
\ee
 Here and in the following, $\bar M$ is a further mass parameter in the problem which sets the mass scale where non-standard kinetic terms become relevant. It probably is of the order of the Planck mass, but we shall keep the general parameter $\bar M$. For this simple class, the speed of sound is, in fact, exactly constant, and not only in lsr. Indeed, for general lagrangians  (\ref{prod-twins}) we find $\mathcal{L}_{XX}\vert = (g_{,XX}/g_{,X})\vert$, which for the higher powers (\ref{hi-pow}) just leads to 
$ \mathcal{L}_{XX}\vert = [(n-1)/X]\vert$.  Expression (\ref{speed-of-sound}), therefore, leads to the speed of sound
\be
c_s^2  = (1+2X\frac{n-1}{X}\vert )^{-1} =(1/(2n-1))
\ee
which respects the bound  (\ref{cs-bound}) for $n\ge 3$. Another example of the product type is
\be
g = \frac{1}{2}X - \frac{\bar M^4}{4}\sin \frac{2X}{\bar M^4}
\ee
leading to (here $\mu \equiv \bar M/M_{\rm P}$, $y \equiv (2/3) (m^2/\mu^4)$)
\be \label{cs-for-sin}
c_s^2 = \left. \left( 1+\frac{4X}{\bar M^4} \frac{\cos\frac{X}{\bar M^4}}{\sin\frac{X}{\bar M^4}} \right)^{-1} \right|
\stackrel{\rm lsr}{=} \frac{1}{1+4\frac{y}{\tan y}} 
\ee
which depends on the two parameters $m$ and $\mu$.  But we know that $m<<1$, so if $\mu \sim 1$ we can take the limit $y\to 0$ and get approximately $c_s \simeq 1/\sqrt{5}$, again within the bound (\ref{cs-bound}).

Next, we consider the class of twins given by a power series in $X$, eq. (\ref{series-twins}), conveniently re-expressed like
\be
\mathcal{L} = \sum_{i=1}^n b_i (\phi) X\left( \frac{X}{\bar M^4}\right)^{i-1} -U(\phi)
\ee
where the $b_i$ are dimensionless functions. In lsr, the first twin condition $\mathcal{L}\vert =1$ simplifies to
\be \label{first-twin-series}
\mathcal{L}\vert \stackrel{\rm lsr}{=} \sum_{i=1}^n ib_i(\phi) y^{i-1} =1
\ee
where $y$ is as above. This condition can be satisfied in many ways, giving rise to different twin models. Obviously, it can also be fulfilled by constant $b_i$. This does {\em not} mean that the $b_i$ of the corresponding twin models can be chosen to be exactly constant, it just means that together with the slow-roll expansion for $W(\phi)$ there exists a slow-roll expansion for the $b_i$ which starts with a constant value, $b_i (\varphi) = (b_{\rm lsr})_i + \mathcal{O} (\varphi^{-2})$ (remember that, for the quadratic potential, the slow-roll expansion is equivalent to an expansion in $\varphi^{-2}$ \cite{infl-adam1}, consistent with the fact that this model is a large-field model where $\varphi_{\rm CMB} >>1$). Higher order slow-roll corrections of the $b_i$ can then be calculated by inserting higher order slow-roll corrections of $W(\phi)$ into the first twin condition (\ref{series-first-cond}). There are other twin models where the $b_i$ are not constant even at lsr, but we will not consider them here. The reason is that for constant $(b_{\rm lsr})_i $, the speed of sound is constant at lsr, too, and reads
\be
c_s^2 \stackrel{\rm lsr}{=} \left( 1 + 2 \sum_{i=1}^n i(i-1) (b_{\rm lsr})_i y^{i-1} \right)^{-1}.
\ee
If we make the additional assumption that all the $b_i$ are non-negative, then this speed of sound is bound by $1\ge c_s \ge 1/\sqrt{2n-1}$. In particular, it takes a value close to $1/\sqrt{2i-1}$ if the term $i(b_{\rm lsr})_i y^{i-1}$ gives the main contribution to the sum (\ref{first-twin-series}).

Finally, we consider the example of dilaton-DBI (=dDBI) as a string- (or SUGRA-)inspired model of inflation. Generically, string theory predicts many scalar fields, but frequently most of them are essentially constant during cosmological inflation, such that cosmological inflation is driven by just one or a few fields (consistent with recent observations, fully compatible with single-field inflation). The DBI lagrangian is \cite{DBI-sky}
\be
\mathcal{L}_{\rm DBI} = -f^{-1}(\phi)\sqrt{1-2Xf(\phi)} + f^{-1}(\phi) - U(\phi).
\ee
The (dimensionless) dilaton field $\chi$ is the Goldstone field of the scale symmetry, which determines its coupling. In particular, it couples to the kinetic term $X$ via $X\to e^{-\chi}X$. We will assume, however, that the dilaton takes a constant value $\chi \to \chi_0$ during inflation, such that its only effect is to change the normalization of the DBI term. Introducing the constant $c=e^{-\chi_0}$ and performing the convenient rescaling $e^{-\chi_0} f \to f$, we arrive at the dDBI lagrangian
\be
\mathcal{L}_{\rm dDBI} = -cf^{-1}(\phi)\sqrt{1-2Xf(\phi)} + cf^{-1}(\phi) - U(\phi).
\ee
The first twin condition $\mathcal{L}_{,X}=1$ leads to $f=[(1-c^2)/(2X\vert )]$, and for the speed of sound we easily find $c_s^2 =c^2 =\;$const. This result follows from the first twin condition alone, so it holds for dDBI twins of arbitrary standard theories (\ref{stand-L}). Returning to the case $V=M^2 \phi^2$, it is obvious that condition (\ref{cs-bound}) can be fulfilled by an appropriate choice of the dilaton field value $\chi_0$.

Finally, the second twin condition (\ref{2-twin}) allows to determine the dDBI potential $U$ in terms of the potential $V$ of the standard twin as
\be
U = V - \frac{1-c}{1+c} X\vert = 6 M_{\rm P}^4 W^2 - \frac{8}{1+c}M_{\rm P}^6 W_{,\phi}^2. 
\ee 
 
%\vspace*{0.4cm}

\section{Summary}
It was the purpose of this paper to demonstrate the simplicity and usefulness of the twin model concept for cosmological inflation. Considering generalized lagrangians $\mathcal{L}(X, \phi)$ is not an arbitrary choice, but enforced on us on theoretical grounds (the corresponding terms are not suppressed near the Planck scale). A formalism which permits to transfer both the classical cosmology and the analytical methods of standard single-field inflation to the realm of K-inflation is, therefore, obviously advantageous. The fact that twins can be found by purely algebraic methods adds to their utility. 
K-field twins can still lead to different predictions for cosmological inflation because of their nontrivial speed of sound. In particular, we found that already for the simplest inflaton theory with a quadratic potential, $V = M^2\phi^2$ (essentially ruled out by recent observations), twins compatible with all observational constraints can be found easily. 
We introduced some simple examples of twins to demonstrate the working and usefulness of the method, but we also considered the eminently physical example of dDBI inflation, which belongs to the most natural lagrangians from the point of view of fundamental theory. Once the dDBI model is expressed as a twin, its predictions for cosmological inflation can be determined with essentially zero additional effort.
Finally, the very general character of the method implies that many more models of physical relevance can be covered by it.

\vspace*{0.2cm}

%%%%%%%%%%%%%%%%%%%%%%%%%%%%%%%%%%%%%%%%%
\section*{Acknowledgements}
%%%%%%%%%%%%%%%%%%%%%%%%%%%%%%%%%%%%%%%%%
%{\em Acknowledgements: \; }
The authors acknowledge financial support from the Ministry of Education, Culture, and Sports, Spain (Grant No. FPA2017-83814-P), the Xunta de Galicia (Grant No. INCITE09.296.035PR and Conselleria de Educacion), the Spanish Consolider-Ingenio 2010 Programme CPAN (CSD2007-00042), Maria de Maetzu Unit of Excellence MDM-2016-0692, and FEDER. 
%\nocite{*}

%%%%%%%%%%%%%%%%%%%%%%%%%%%
\appendix
\section{The twin conditions in the Hamilton-Jacobi formalism}
%%%%%%%%%%%%%%%%%%%%%%%
The general idea of this approach is to express the homogeneous (spatially constant) systems resulting from $\bar{\mathcal{L}}$ and $\mathcal{L}(X, \phi)$, relevant for the inflationary evolution, by  equivalent  mechanical systems, and to derive their corresponding Hamilton-Jacobi (H-J) equations in both cases. The resulting H-J equations for the two systems are different, but we may impose the condition that these two H-J equations have a common solution which describes the (same) inflationary evolution. We shall see that the consistency of this assumption requires to impose the two twin conditions on the generalized Lagrangian $\mathcal{L}(X, \phi)$.

We start from the full action
\begin{equation}
S = S_{\rm EH} + S_{\rm m} = \int d^4 x \sqrt{|g|} ( -\frac{M_{\rm P}^2}{2} R + \mathcal{L}_{\rm m} )
\end{equation}
where $ S_{\rm EH} $ is the Einstein-Hilbert (EH) action and $S_{\rm m}$ the matter (inflaton) action. Assuming the FRW line element (\ref{FRW-metric}) and an inflaton field which only depends on time, and ignoring the space dependence ($d^4 x \to dt$) we arrive at an equivalent mechanical system with the two generalized coordinates $a(t)$ and $\phi (t)$.  The EH action produces a term with second derivatives because $R = -6 [(\dot a  ^2+a \ddot a)/a^2]$, but the second derivative may be eliminated by a partial integration, $\int dt a^3 R = -6\int dt\, a (\dot a^2 + a\ddot a) = 6 \int dt \, a \dot a^2 + \rm{b.c.}$ where we shall ignore the boundary contribution b.c. The resulting mechanical (total) Lagrangian is
\be
\mathcal{L}_{\rm t} = -3M_{\rm P}^2 a\dot a^2 + a^3 \mathcal{L}_{\rm m}.
\ee
For a standard Lagrangian $\mathcal{L}_{\rm m} = \bar{\mathcal{L}}$ we get
\be
\bar{\mathcal{L}}_{\rm t} = -3M_{\rm P}^2 a\dot a^2 + a^3 \left( \frac{1}{2}\dot\phi^2 - V(\phi) \right) ,
\ee
leading to the generalized momenta
\be
\pi_a = \frac{\partial \bar{\mathcal{L}}_{\rm t}}{\partial \dot a} = -6M_{\rm P}^2 a\dot a \; ,\quad 
\pi_\phi = \frac{\partial \bar{\mathcal{L}}_{\rm t}}{\partial \dot \phi} = a^3 \dot \phi
\ee
and to the Hamiltonian
\bea 
\bar{\mathcal{H}}_{\rm t} &=& \pi_a \dot a + \pi_\phi \dot \phi - \bar{\mathcal{L}}_{\rm t} = -3M_{\rm P}^2 a\dot a^2 + \frac{1}{2} a^3 \dot \phi^2 + a^3 V \nonumber \\
&=& -\frac{1}{12 M_{\rm P}^2 a}\pi_a^2 + \frac{1}{2a^3} \pi_\phi^2 + a^3 V \\ &\equiv & \bar{\mathcal{H}}(a, \phi ,\pi_a ,\pi_\phi ) .
\eea
The Hamilton-Jacobi (H-J) formalism now consists in assuming a generating function $F(q_i, P_i)$ for a canonical transformation of the {\em second} type (i.e., depending on the old coordinates $q_i$ and new momenta $P_i$) to a new system of coordinates $Q_i$ and momenta $P_i$ such that the new coordinates $Q_i$ are cyclic and the new momenta $P_i$, are constants of motion, $P_i =$ const. The old momenta $p_i$ are then no longer independent variables but, instead, given by 
$p_i = [(\partial F)/(\partial q_i)]$. Finally, the assumption that all new coordinates are cyclic implies that the new Hamiltonian is identically zero, which is expressed by the H-J equation (here $\mathcal{H}$ is the old Hamiltonian) 
\be
\mathcal{H}\left( q_i , [(\partial F)/(\partial q_i)]\right) + [(\partial F)/(\partial t)] =0.
\ee
$F$ is called Hamilton's principal function in this context. 
Here, arbitrary solutions for the original mechanical system are provided by the general solutions $F$ of this first-order PDE, and the corresponding arbitrary values for the conserved new momenta are related to the arbitrary integration constants on which the solutions $F$ depend. 
 
In our case, however, we are not interested in the most general solution $\bar F(a,\phi ,t)$ but in the particular solution which is supposed to describe the homogeneous inflationary evolution. Making, in a first step, the simplifying assumption of a time independent $\bar F$,  $[(\partial \bar F)/(\partial t)] =0$, we get
\bea
\bar{\mathcal{H}}_{\rm t}(a, \phi ,\partial \bar F/\partial a ,\partial \bar F/ \partial \phi ) &=& \nonumber \\
 -\frac{1}{12M_{\rm P}^2 a} (\bar F_{,a})^2 + \frac{1}{2a^3 }(\bar F_{,\phi})^2 + a^3 V &=& 0.
\eea
Upon inspection, it is easy to guess the further simplifying assumption
\be
\bar F (a,\phi ) = a^3 \bar G(\phi ),
\ee
leading to
\be
-\frac{3}{4M_{\rm P}^2} \bar G^2 + \frac{1}{2}\bar G_{,\phi}^2 + V =0
\ee
which, after the identification $\bar G(\phi) = -2\sqrt{2}M_{\rm P}^3 W(\phi)$, is identical to the superpotential equation (\ref{superpot-eq}). 
This is a first order ODE (ordinary differential equation) with a one-parameter family of solutions. For slow-roll inflation, which is the case we explicitly consider in this paper, there exists the so-called attractor solution among these solutions. All other solutions quickly converge to this attractor, so that $W(\phi)$ at the end of inflation is essentially unique (see appendix B). 
%It is one of the attractive features of the superpotential equation (as opposed to the FRW equations) that for a wide class of potentials the attractor solution can be found exactly \cite{infl-adam1}. 
Outside the range of slow-roll inflation, more careful considerations are required to select the relevant inflationary solution from this one-parameter family. The issue of inflationary scenarios beyond the paradigm of slow-roll lies, however, beyond the scope of the present paper, although the H-J formalism and the twin model concept may be applied equally well to non-slow-roll models of inflation.

Finally, using $\pi_a = \bar F_{,a}$ and $\pi_\phi = \bar F_{,\phi}$, we get for the velocities
\be
\dot a = -\frac{a}{2M_{\rm P}^2} \bar G \; \Rightarrow \; H = -\frac{1}{2M_{\rm P}^2} \bar G
\ee 
and
\be \label{stan-root-phi}
\dot \phi = \bar G_{,\phi}.
\ee

For a generalized Lagrangian $\mathcal{L}_{\rm m} = \mathcal{L}(X, \phi)$ (where now $X \equiv (1/2) \dot \phi^2$) we have
\be
\mathcal{L}_{\rm t} = -3M_{\rm P}^2 a\dot a^2 + a^3  \mathcal{L}(X, \phi),
\ee
leading to the generalized momenta
\be
\pi_a = \frac{\partial \mathcal{L}_{\rm t}}{\partial \dot a} = -6M_{\rm P}^2 a\dot a 
\ee
like in the standard case, and
\be
\pi_\phi = \frac{\partial \mathcal{L}_{\rm t}}{\partial \dot \phi} = a^3 \mathcal{L}_{,\dot \phi}
= a^3 \dot \phi \mathcal{L}_{,X}.
\ee
Inverting this last equation, i.e., expressing $\dot \phi$ in terms of $\pi_\phi$, $\phi$ and $a$ is, in general, complicated and may lead to several roots, 
\be \label{root}
\dot \phi = f^{(i)} (\phi ,\pi_\phi /a^3)
\ee
where the possible roots are indicated by the index $(i)$. For our purposes, however, we do not need the explicit expression of the relevant root, its existence is sufficient. For the Hamiltonian we get (using $\dot \phi \mathcal{L}_{,\dot \phi} = 2X \mathcal{L}_{,X}$)
 \bea 
\mathcal{H}_{\rm t} &=& \pi_a \dot a + \pi_\phi \dot \phi - \mathcal{L}_{\rm t} 
\nonumber \\
&=& -3M_{\rm P}^2 a\dot a^2 +  a^3 \left( 2X\mathcal{L}_{,X} - \mathcal{L} \right) \nonumber \\
&=& -\frac{1}{12 M_{\rm P}^2 a}\pi_a^2 +  a^3 K(\phi ,\pi_{\phi}/a^3) \\ &\equiv & \mathcal{H}_{\rm t}(a, \phi ,\pi_a ,\pi_\phi ) ,
\eea
where
\be
K(\phi ,\pi_{\phi}/a^3) \equiv \left. \left( 2X\mathcal{L}_{,X} - \mathcal{L} \right) \right|_{\dot \phi = f^{(i)}}
\ee
is the expression which results when $2X\mathcal{L}_{,X} - \mathcal{L}$ is evaluated on a particular root (\ref{root}) of the velocity. The resulting H-J equation, for a time-independent principal function $F$, reads 
\be
-\frac{1}{12 M_{\rm P}^2 a}(F_{,a})^2 +  a^3 K(\phi ,F_{,\phi}/a^3) =0
\ee
and, after the obvious simplifying ansatz $F(a,\phi) = a^3 G(\phi)$ leads to
 \be
 -\frac{3}{4M_{\rm P}^2}G^2 +\left.  \left(2X\mathcal{L}_{,X} - \mathcal{L}\right) \right| =0
 \ee
 where now the vertical line stands for the evaluation of the velocity  $\dot \phi $ at
 \be
 \dot \phi = f^{(i)}(\phi ,G_{,\phi}).
 \ee
 $a$ still obeys
 \be
\dot a = -\frac{a}{2M_{\rm P}^2}  G \; \Rightarrow \; H = -\frac{1}{2M_{\rm P}^2}  G.
\ee 
Now we assume that $\mathcal{L}$ is a twin of $\bar{\mathcal{L}}$, sharing the same inflationary solution $\phi(t)$, $a(t)$ with the same matter energy expression $\rho (t) = \bar \rho (t)$. Equality of the inflaton fields implies
\be
\dot \phi = f^{(i)}(\phi ,G_{,\phi}) = \bar G_{,\phi},
\ee
i.e., one of the roots $f^{(i)}(\phi ,G_{,\phi})$ must coincide with the standard root (\ref{stan-root-phi}). 
But this condition is precisely the "on-shell evaluation" condition of section II.
Next, the equality of the cosmic scale factors $a(t)$ implies
\be
G(\phi )=\bar G(\phi),
\ee
i.e., the two principal functions $F$ and $\bar F$ must coincide. This  implies that the on-shell values of the momenta $\pi_\phi$ coincide, which after a division by $a^3$ leads to
\be \label{HJ-twin1}
\bar G_{,\phi} = \dot \phi \vert = \dot \phi \mathcal{L}_{,X}\vert \; \; \Rightarrow \; \;
\mathcal{L}_{,X}\vert =1,
\ee
i.e., precisely the first twin condition of section II. Finally, equality of the energies (corresponding to the matter energy densities of the original field theories) implies
\be \label{HJ-twin2}
\bar \rho = (2X -\bar{\mathcal{L}})\vert = \rho = (2X \mathcal{L}_{,X} -\mathcal{L} )\vert \; \Rightarrow \; 
\mathcal{L}\vert = \bar{\mathcal{L}}\vert ,
\ee
i.e., the second twin condition of section II.

To summarize, the assumption that a standard inflaton theory $\bar{\mathcal{L}}$ shares a common (inflationary) solution $a(t)$, $\phi(t)$ with coinciding energy densities with a generalized theory $\mathcal{L}(X, \phi)$ implies that their corresponding H-J equations must share a common solution, too.
Further, this assumption is consistent only provided that the non-standard twin obeys the two algebraic consistency conditions ("twin conditions") (\ref{HJ-twin1}) and (\ref{HJ-twin2}). 
We remark that, while the two H-J equations share a common solution, they are completely different equations, in general. The corresponding field theories are, therefore, {\em not} related by a field redefinition or, for the equivalent mechanical systems, by a canonical transformation.

\section{The attractor in the superpotential formalism}
It is well-known that for the standard FRW equations, for the case of slow-roll inflation there exists a particular attractor solution to which all other inflationary solutions quickly converge, see e.g.,  \cite{Muk}. Here we want to show that this attractor solution continues to exist in the superpotential formalism, as must, of course, be true. 
Indeed, the superpotential equation may be expressed like
\be
W_{,\phi} = \pm \frac{1}{2M_{\rm P}^3} \sqrt{6M_{\rm P}^4 W^2 - V}
\ee
where the $+$ sign corresponds to $\dot \phi <0$, and the minus sign to $\dot \phi >0$, see eq. (\ref{dot-H}). For concreteness we assume the $-$ sign, the other case can be treated analogously.
Now assume that we choose two different initial values $W_{1,i} \, > \, W_{2,i}$ at some initial inflaton field value $\phi = \phi_i$ where inflation (or its cosmologically relevant phase) begins, giving rise to two different solutions $W_1 (\phi)$ and $W_2 (\phi)$ during inflation (for $\phi_i < \phi < \phi_e$). It is obvious from the above equation that $W_1 (\phi ) > W_2 (\phi)$ implies $W_{1,\phi} (\phi ) < W_{2,\phi} (\phi)$. On the other hand, the inequality $W_1 (\phi ) > W_2 (\phi)$ is maintained during inflation, because different trajectories from a one-parameter family of solutions can never intersect or touch.
This implies that the two solutions $W_1 (\phi)$ and $W_2 (\phi)$ approach each other. As the argument holds for two arbitrary solutions (two arbitrary initial values $W_{1,i} $ and $ W_{2,i}$), this further implies that arbitrary solutions converge to each other and, therefore, towards a common attractor solution. Numerical calculations show that the convergence is, in fact, very fast \cite{infl-adam1}.

\bibliographystyle{apsrev4-1}
%%\bibliography{refs}

\begin{thebibliography}{16}
\bb{starob1}
A. Starobinsky, Phys. Lett. B91 (1980) 99.
\bb{guth1}
A. Guth, Phys. Rev. D23 (1981) 347.
\bb{linde1}
A. Linde, Phys. Lett. B108 (1982) 389.
\bb{liddle-lyth}
A.R. Liddle, D.H. Lyth, {\it Cosmological Inflation and Large-Scale Structure},
Cambridge University Press, Cambridge, 2000.
\bb{Dod}
S. Dodelson, {\it Modern Cosmology}, Academic Press, New York, 2003.
\bb{Muk} V. Mukhanov, {\it Physical Foundations of Cosmology}, Cambridge University Press, Cambridge, 2005.
\bibitem{Weinberg} S. Weinberg, {\it Cosmology \/},  Oxford University Press, New York, 2008.
\bibitem{Baumann}D. Baumann (2009), {\it TASI Lectures on Inflation \/} \href{https://arxiv.org/pdf/0907.5424.pdf}{(arXiv:0907.5424)}
\bb{WMAP1}
WMAP Collaboration,
The Astrophysical Journal Supplement Series, 208, 19, 2013; arXiv:1212.5226. 
\bb{WMAP2}
WMAP Collaboration,
The Astrophysical Journal Supplement Series, 208, 20, 2013; arXiv:1212.5225. 
\bb{Planck-infl-2013}
Planck Collaboration XXII, Planck 2013 results XXII, Constraints on Inflation, Astron. Astrophys. 571, A22 (2014); arXiv:1303.5082.
\bb{Planck-infl-2015}
Planck Collaboration XX, Planck 2015 results XX, Constraints on Inflation, Astron. Astrophys. 594, A20 (2016); arXiv:1502.02114.
\bb{Planck}
Planck Collaboration X, Planck 2018 results X, Constraints on inflation, arXiv:1807.06211.

\bb{enzy}
J. Martin, C. Ringeval, V. Vennin, Phys. Dark Univ. 5-6 (2014) 75.

\bb{k-infl}
C. Armendariz-Picon, T. Damour, V. Mukhanov, Phys. Lett. B458 (1999) 209.
\bb{k-infl2}
J. Garriga, V. Mukhanov, Phys. Lett. B458 (1999) 219.
\bb{SeLi1}
%Primordial non-Gaussianities in single field inflation
D. Seery, J.E. Lidsey,
JCAP 0506 (2005) 003.
\bb{Kachru2007}
X. Chen, M.-X. Huang, S. Kachru, G. Shiu, JCAP 0701 (2007) 002; arXiv: hep-th/0605045.
\bb{Vennin3}
%K-inflationary Power Spectra at Second Order
J. Martin, C. Ringeval, V. Vennin,
JCAP 1306 (2013) 021.
\bb{Twin-models}
Twin models of a scalar field theory without gravity were first introduced in \cite{twins-trodden} and further studied in \cite{twins-baz1}, \cite{twins-baz2}, \cite{twins-zhong1}, both in flat space and in a brane-world context. The algebraic twin conditions were derived in \cite{twins-adam1}, \cite{twins-adam2}. Finally, twin models in a cosmological context were first studied in \cite{twins-baz3}, \cite{twins-zhong2}, and used to produce some exact solutions.
\bb{twins-trodden} M. Andrews, M. Lewandowski, M. Trodden, D. Wesley, Phys. Rev. D82 (2010) 105006; arXiv:1007.3438.
\bb{twins-baz1} 
D. Bazeia, J.D. Dantas, A.R. Gomes, L. Losano, R. Menezes,
Phys. Rev. D84 (2011) 045010; arXiv:1105.5111.
\bb{twins-baz2} 
D. Bazeia, R. Menezes, Phys. Rev. D84 (2011) 125018;
arXiv:1111.1318.
\bb{twins-zhong1}
Yuan Zhong, Yu-Xiao,
Class. Quant. Grav. 32 (2015) no.16, 165002;
arXiv:1408.6416.
\bb{twins-adam1}
C. Adam, J.M. Queiruga, Phys. Rev. D84 (2011) 105028;
arXiv:1109.4159.
\bb{twins-adam2}
C. Adam, J.M. Queiruga,
Phys. Rev. D85 (2012) 025019;
arXiv:1112.0328.
\bb{twins-baz3}
D. Bazeia, J.D. Dantas,
Phys. Rev. D85 (2012) 067303;
arXiv:1202.5978. 
\bb{twins-zhong2}
Yuan Zhong, Chun-E Fu, Yu-Xiao Liu,
Sci. China Phys. Mech. Astron. 61 (2018) no.9, 90411;
arXiv:1604.06857.

\bb{bond1990}
D.S. Salopek, J.R. Bond, Phys. Rev. D {\bf 42} (1990) 3936.
\bb{lyth1993}
E.D. Stewart, D.H. Lyth,
Phys. Lett. B302 (1993) 171;
gr-qc/9302019. 
\bb{kinney1997}
W.H. Kinney, Phys. Rev. D {\bf 56} (1997) 2002.

\bb{nunez2004}
D.Z. Freedman, C. Nunez, M. Schnabl, K. Skenderis, Phys. Rev. D {\bf 69} (2004) 104027.
\bb{skenderis2006}
K. Skenderis, P.K. Townsend, Phys. Rev. Lett. {\bf 96} (2006) 191301.
\bb{garriga2016}
%%delta-N formalism from superpotential and holography
J. Garriga, Y. Urakawa, F. Vernizzi,
JCAP {\bf 1602} (2016) 036.
\bb{bazeia1}
%% First-order formalism and dark energy
D. Bazeia, C.B. Gomes, L. Losano, R. Menezes,
Phys. Lett. B {\bf 633} (2006) 415.
\bb{bazeia3}
%%First-order formalism for dark energy and dust
D. Bazeia, L. Losano, J.J. Rodrigues, R. Rosenfeld,
Eur. Phys. J. C {\bf 55} (2008) 113.
\bb{kir1}	
%Asymptotic freedom, asymptotic flatness and cosmology 
E. Kiritsis, JCAP 1311 (2013) 011.
\bb{kir2}
%Universality classes for models of inflation 
P. Binetruy, E. Kiritsis, J. Mabillard, M. Pieroni, C. Rosset,
JCAP 1504 (2015) 033. 
\bb{FOEL}
C. Adam, F. Santamaria,
JHEP 1612 (2016) 047. 

\bb{infl-adam1}
C. Adam, D. Varela, arXiv:1901.02892.

\bibitem{Schwarz}D.J. Schwarz, C. Terrero-Escalante, A. Garcia, Phys. Lett. B 517 (2001) 243.
\bb{Vennin2}
V. Vennin, Phys. Rev. D89 (2014) 083526.
\bb{footnote}
In the standard FRW treatment of cosmological inflation, the leading slow-roll approximation is achieved by neglecting the term $X=(1/2)\dot \phi^2$ in comparison to the potential $V$ in the FRW equation (\ref{FRW-eq1}) for a standard energy density $\bar \rho = X+V$, leading to the lsr approximation $H=\sqrt{V/3M_{\rm P}^2}$, which is completely equivalent to $W_{\rm lsr} =M_{\rm P}^{-2}\sqrt{V/6}$.
\bb{DBI-sky}
%DBI in the sky
M. Alishahiha, E. Silverstein, D. Tong,
Phys. Rev. D70 (2004) 123505.



\end{thebibliography}
%

\end{document}